\documentclass[twocolumn,floatfix,aps, prx, superscriptaddress]{revtex4-2}

\usepackage{graphicx}
\usepackage{mathtools}
\usepackage{amsmath}
\usepackage{amssymb}
\usepackage{bm}
\usepackage{color}
\usepackage{comment}
\usepackage[]{footmisc}
\usepackage{lineno}
\usepackage{dsfont}
\usepackage{soul,xcolor}
\usepackage[pdfborder={0 0 0}]{hyperref}
\usepackage{nicematrix}
\usepackage{subcaption}
\usepackage{tabularray}
\usepackage{physics}
\usepackage{pgfplots}
\pgfplotsset{width=10cm,compat=1.9}
\usepackage{ragged2e} 
\usepackage{ulem}

\usepackage{graphicx,overpic}

\newcommand{\clb}{\color{blue}}

\renewcommand{\emph}[1]{{\it #1}}

\begin{document}

\title{Step-Edge Anomaly in Topological Metals} 

\author{O.~Schweizer}
\author{V.~Gali}
\affiliation{Dahlem Center for Complex Quantum Systems, Halle-Berlin-Regensburg Cluster of Excellence CCE, and Fachbereich Physik, Freie Universit\"at Berlin, Arnimallee 14, 14195 Berlin}

\author{A.Y.~Chaou}
\affiliation{Dahlem Center for Complex Quantum Systems, Halle-Berlin-Regensburg Cluster of Excellence CCE, and Fachbereich Physik, Freie Universit\"at Berlin, Arnimallee 14, 14195 Berlin}
\affiliation{Donostia International Physics Center, P. Manuel de Lardizabal 4, 20018 Donostia-San Sebastian, Spain}

\author{G.~Lemut}
\author{P.W.~Brouwer}
\author{M.~Breitkreiz}
\email{breitkr@physik.fu-berlin.de}
\affiliation{Dahlem Center for Complex Quantum Systems, Halle-Berlin-Regensburg Cluster of Excellence CCE, and Fachbereich Physik, Freie Universit\"at Berlin, Arnimallee 14, 14195 Berlin}
 
\begin{abstract}
 Bulk–boundary correspondence guarantees the presence of robust, anomalous states on the boundary of topological matter. The edges of a two-dimensional Chern insulator harbor one-dimensional chiral states, which have a conductance $n\, e^2/h$, where $n$ is an integer that is solely determined by the bulk. In this work we show that step edges on the surface of three-dimensional topological metals have a robust  conductance  $K\, e^2/h$, where $K$ is also fixed by the bulk and assumes non-integer values. We explain this prediction on the basis of the topology of gapless systems, exemplify it on a lattice model, and connect to recent experimental observations of enhanced density of states at step edges in topological metals.
\end{abstract}

\maketitle

\emph{\clb Introduction}---Bulk–boundary correspondence is a unique manifestation of the topology of a bulk material in form of anomalous states on its surface, where the attribute \emph{anomalous} expresses that the surface states cannot exist independently from the higher-dimensional topological bulk \cite{RevModPhys.82.3045,RevModPhys.83.1057,Trifunovic2019,Armitage2017}, which, in turn, guarantees the presence and robustness of anomalous surfaces states given a topological bulk system. This link underpins robust transport properties of the surface, forming the foundation for applications such as in low-dissipation information processing \cite{Gilbert2025,Kiani2025}, and enables the identification of bulk topology from surface phenomena \cite{Lv2015,Xu2015}.
States bound to defects \cite{Ran2009,Teo2010,deJuan2014,Bulmash2015,Queiroz2024} can also play this role if the bulk topology is of the ``weak'' type, but without weak topology, defect-bound states are often only suggestive, rather than implicative, of a topological bulk.

The discovery of novel, unambiguous manifestations of topology on the material surfaces is of great interest as it can substantially expand the application landscape of topological materials as well es tools for identification. Here we predict such an anomalous manifestation at step edges on otherwise featureless surfaces of three-dimensional topological metals, motivated by experimental observations of enhanced and continuous density of states there \cite{Dong2019,Howard2021,Nag2026}.
In topological metals \cite{Armitage2017,Yan2017,Burkov2017}, the bulk is characterized by topologically protected band touching points that resemble the dispersion of Weyl fermions and its various generalizations and come in pairs of opposite chirality. 
The corresponding boundary states are open sheets of surface states connecting the projections of opposite-chirality Weyl cones onto the surface Brillouin zone (BZ). The open equi-energy contours at the Fermi energy are called Fermi arcs. On surfaces for which the projections of opposite-chirality Weyl nodes overlap, Fermi arcs are absent.

A surface with a Fermi arc has an anomaly that can be quantified in terms of a conductivity. Let the pair of Weyl nodes be separated  by $2\pi \bm{K}$. The Fermi-arc surface band are chiral surface modes moving in the direction $\bm{K}\times\bm{e}_\mathrm{s}$, where $\bm{e}_\mathrm{s}$ is the surface normal, for each surface momentum component along $(\bm{K}\times\bm{e}_\mathrm{s})\times \bm{e}_\mathrm{s}$ between the projected nodes.
Application of a uniform scalar potential $e\,dV$ at the surface induces the charge-current density $d\bm{j} = (e^2/h) \ \bm{K}\times\bm{e}_\mathrm{s}\, dV$, where $e^2/h$ is the unit of conductance. The resulting conductivity quantifying the surface anomaly, 
\begin{equation}
\sigma_\mathrm{s} = \frac{e^2}{h} \,  |\bm{K}\times\bm{e}_\mathrm{s} |,
\label{ifa}
\end{equation}
depends only on the separation of the Weyl nodes in the bulk band structure and the surface orientation and is independent of the detailed surface termination. It manifests experimentally as an anomalous Hall conductivity \cite{Burkov2011,Suzuki2016} in time-reversal broken systems, such as the considered minimal model of two Weyl nodes. In time-reversal-preserved topological metals, Fermi arcs come in countermoving pairs, in which case the longitudinal conductivity contribution of Fermi arcs manifests  as inverse resistance scaling \cite{Zhang2019,Breitkreiz2019,Piskunow2021,Khan2025,Lanzillo2024,Lien2023,Cheon2025}. 

\begin{figure}[t]
    \includegraphics[width=\columnwidth]{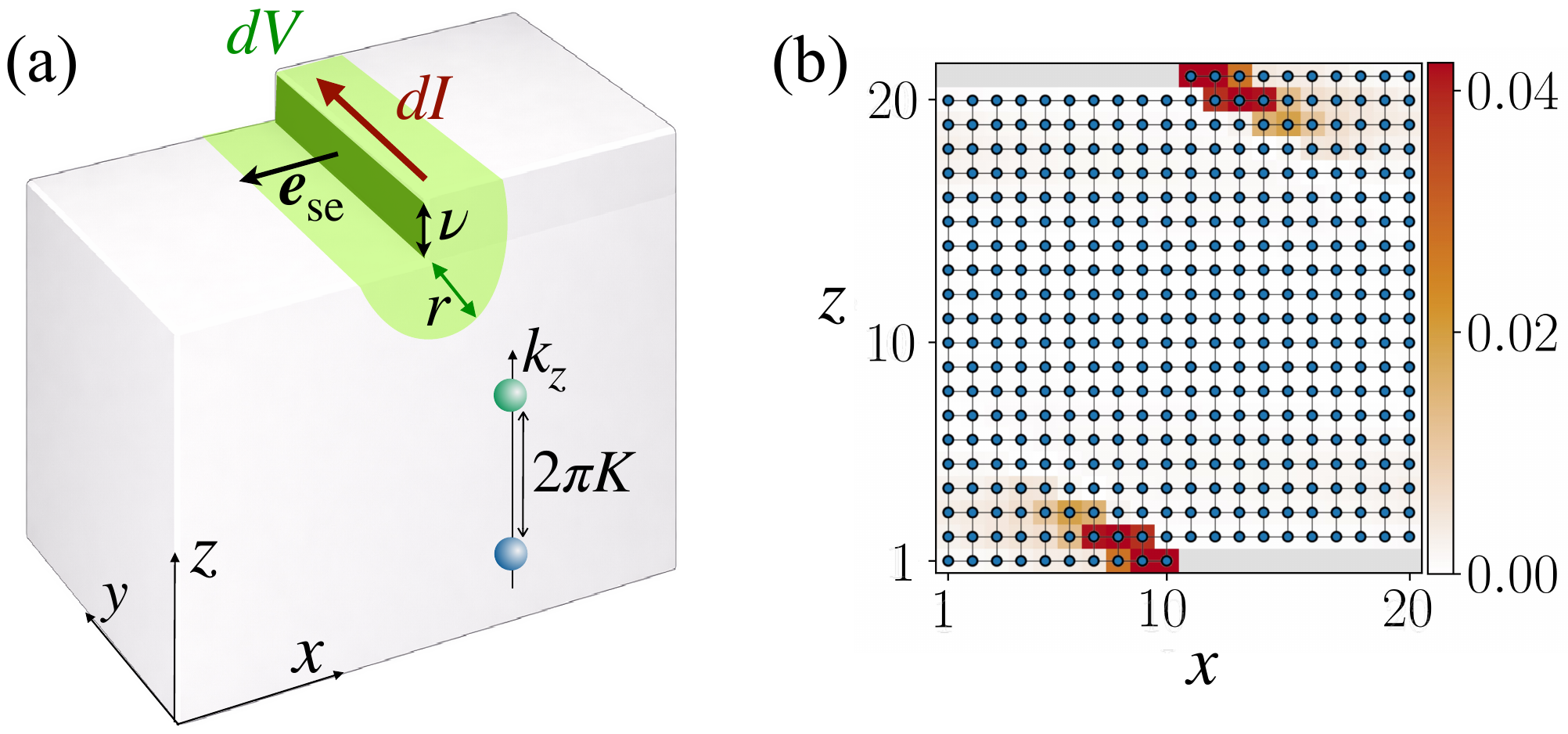}
     \caption{\justifying (a) Weyl semimetal with two Weyl nodes separated by $2\pi K$ in the $k_z$ direction. The upper surface consists of two half-planes at constant $z$, separated by a step edge of height $\nu$. A local potential $dV$ applied at the step edge results in a current response $dI$ in the $y$ direction. (b) Lattice geometry in $x$- and $z$-direction with step edges on top and bottom surfaces for system size $L_x = L_z = 20$ with overlay of local density of states at $\varepsilon=0.1$ for Weyl node separation of $K=5/6$. At the top and bottom step edges two states localize. }
    \label{fig:1}
\end{figure}

In this work, we extend this bulk-surface correspondence to a bulk-step-edge correspondence for a step edge in which the surface is shifted by an integer number of unit cells \footnote{Localized states at step
edges with a height less than one unit cell were observed by \cite{Howard2021}. We do
not consider such states here, since they depend on the topology of a
sub-monolayer at the surface and not on topology of the bulk crystal}. In particular, we show that a pair of Weyl nodes with separation $2\pi \bm{K}$ gives rise to an anomalous step-edge conductance 
\begin{equation}
	G_\mathrm{se} =  \frac{e^2}{h} \nu\, |\bm{K}\times \bm{e}_\mathrm{se} | 
	\label{ise}
\end{equation}
where $\bm{e}_\mathrm{se} $ is the unit vector normal to the mini-surface of the step edge and $\nu$ is its height, see Fig.\ \ref{fig:1}(a). The conductance $ G_\mathrm{se}$ can assume a non-integer multiple ($\nu |\bm{K}\times \bm{e}_\mathrm{se} |$) of the unit of conductance. This is different from the anomalous conductance associated with discrete chiral modes, such as in a Chern insulator or a finite Weyl-semimetal surface \cite{Burkov2011} (see below), which is quantized in integer units of $e^2/h$. Here we instead show that 
the fractional response of Eq.\ (2) is a combination of the quantized response of modes localized at the step edge and a non-quantized local response carried by bulk modes. The number of localized modes at the step edge (if any) depends on local details at the step edge. However, the value of the conductance does not. Note that Eq.\ \eqref{ise} is nothing but the application of Eq.\ \eqref{ifa} to the mini-surface of the step edge. 
However, the total conductance associated with the surface anomaly of Eq.\ \eqref{ifa} is subject to finite-size corrections, which, e.g., force the net conductance at a finite surface to be quantized \cite{Burkov2011}. If applied to a surface with a width of a few unit cells, finite-size corrections would be of the same order as the conductance itself. Our prediction is that such corrections are \emph{absent} for the mini-surface of a step edge.

Before we detail the mechanism behind the non-integer step-edge conductance, we first use a simple \emph{Gedankenexperiment} \cite{Nag2026} to explain its origin in the bulk band structure of the topological semimetal. 
We consider a model system of Fig.\ \ref{fig:1}(a), but first with a flat upper surface without a step edge. Consider that this upper surface is in the plane that is slightly tilted by an angle $\theta$ around the $y$ axis from the constant-$z$-plane. The Weyl node projections onto the new surface BZ will be separated by $K\sin(\theta)$ and the resulting small Fermi arc carries chiral current density $dj_\mathrm{t} =dV\,(e^2/h) K\sin(\theta)$. Since this current only depends on the bulk and the global surface orientation, it is invariant with respect to microscopic modifications of the surface. The tilted surface may thus consist of a
periodic sequence of large constant-$z$ plateaus of size $L$ and small steps of height $\nu$ with $\nu/L = \tan \theta$.
The current per step edge is thus $dI_\mathrm{se}= (dj_\mathrm{t})\times L/\cos(\theta)  = dV (e^2/h)\nu K$. Since the large constant-$z$ plateau is a surface onto which Weyl nodes project onto one another it does not support a chiral current, so that $dI_\mathrm{se}$ must be concentrated at the step edge. The step-edge conductance $dI_\mathrm{se}/dV$ is thus as given in Eq.\ \eqref{ise}.

While this argument is appealing because of its simplicity, some questions remain: The tilted-surface argument does not resolve contributions of eventual localized modes at the step edge and bulk modes and it does not address the energy window within which the universal step-edge conductance applies. The size of the energy window matters, because for a tilted surface the Fermi-arc length and, hence, the energy window in which well-defined surface modes exist goes to zero as the tilt angle $\theta \to 0$.
In the calculations below we address these questions by first considering a numerical simulation on a minimal lattice model, followed by an analytical calculation.  

\emph{\clb Lattice simulation}---As a simple model realization of a system exhibiting the step-edge conductance \eqref{ise}, we consider a tight-binding model whose low-energy physics is governed by two Weyl cones of opposite chirality described by the two-band Hamiltonian 
\begin{align}H &= [2 - \cos(k_y)-\cos(k_x) + \cos(\pi K)-\cos(k_z)]\, \sigma_z  \nonumber\\
&+ \sin(k_x) \, \sigma_x + \sin(k_y) \, \sigma_y\,, 
\label{eqha}
\end{align}
where $\{k_i\}$ are momenta, $\{\sigma_i\}$ denote Pauli matrices,  and lattice constants, hopping amplitudes, and $\hbar$ are set to one. The Weyl nodes are located at $k_z=\pm \pi K$. We consider the model \eqref{eqha} on a slab geometry with surfaces perpendicular to the $z$ axis and thickness $L_z$. Since the projections of the Weyl points onto the surface Brillouin zone coincide, such a surface has no Fermi arc state.

To identify the current carried at step edges, we consider step edges of unit height, running along the $y$ direction, on the top and bottom surfaces. Since the step edges run in the $y$ direction, our system is manifestly translation invariant along $y$.  To avoid Fermi arc states at open boundaries at $x=0,L_x$, we extend the system periodically along $x$. The period is $L_x$ plus a shift by $\nu$ unit cells along $z$, see Fig.\ \ref{fig:1}(b) for $\nu=1$. Although $L_x$ and $L_z$ must be finite in our numerical calculations, we take them large enough that our results do not depend on them.

\begin{figure}[t]
    \centering
    \includegraphics[width=\linewidth]{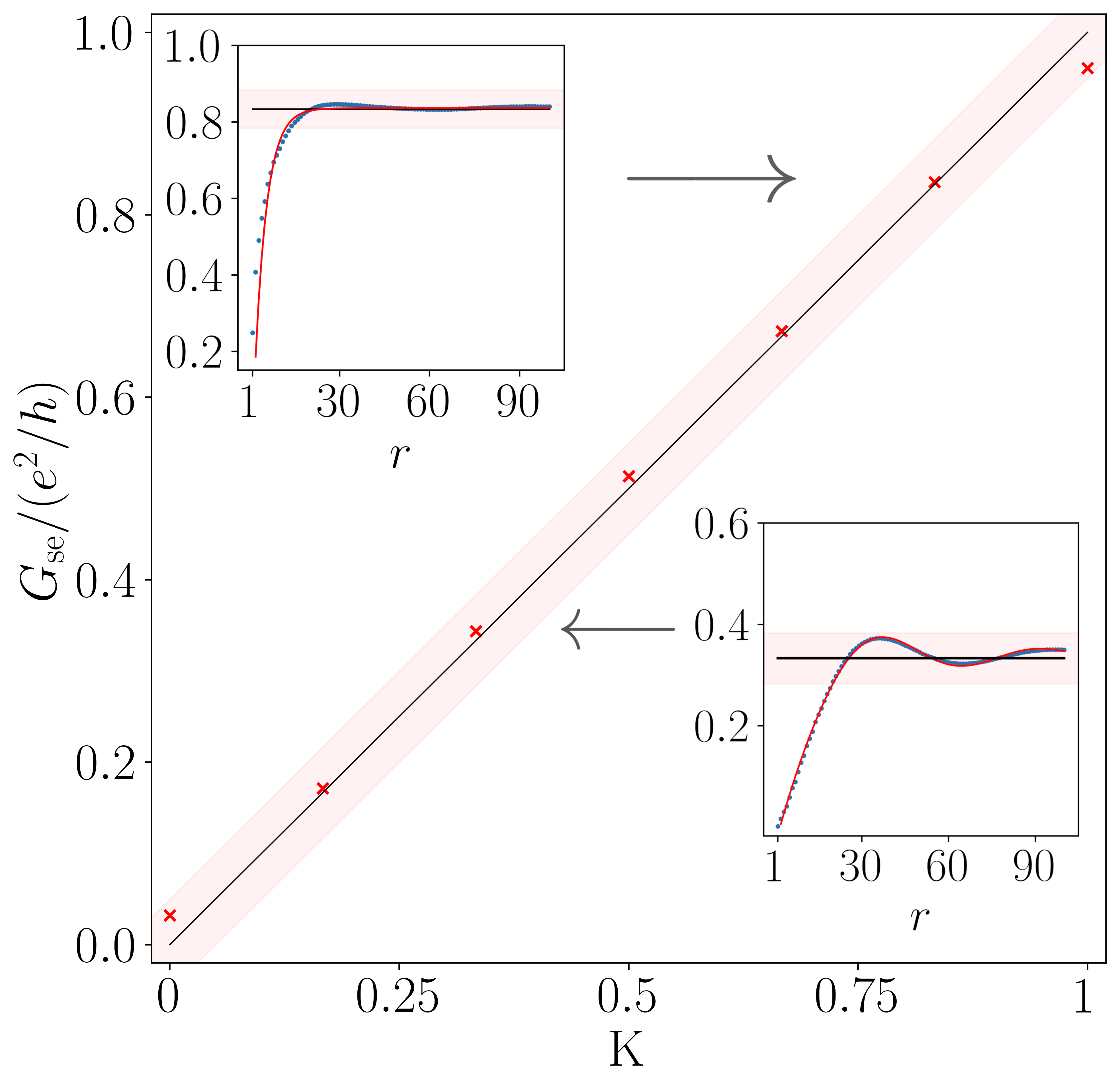}
    \caption{\justifying Step edge conductance (red crosses) as defined in \eqref{eq.4} at different values of the Weyl-node separation $2 \pi K$ for system size of $L_x=L_z=400$, step edge of height $\nu=1$, and energy $\varepsilon=0.05$. The black line is the universal conductance of Eq.~\eqref{ise}. Insets show step edge conductance for $K=1/3$ and $K=5/6$ as a function of radius $r$ of the half-cylindrical region in which the potential $dV$ is applied. The red shaded region is a window of width $0.1$ to visualize scales between the insets and the main plot. At each $K$, the numerically calculated response is fitted to a function of the form $f(r) = A(1 - e^{-B r}) + C \sin(D(r-{\bar r}))/(r - {\bar r})$ to extract the value $A$ of the step edge conductance (for more details, see SM \cite{SM}).}
    \label{fig:2}
\end{figure}

We compute the eigenstates of the lattice model at a fixed energy $\varepsilon$ for all $k_y$  using the transfer matrix of a finite sized layer (see SM \cite{SM} for more details). The local density of states shows enhanced weight at both step edges, see Fig.\ \ref{fig:1}(b).
We calculate the step-edge conductance $G_{\mathrm{se}}=d I/d V$ by calculating the  current $d I$  in the $y$ direction in linear 
response to a voltage $d V$ applied to a half-cylindrical region within a distance $\le r$ from the step edge. The current is obtained by summing the $y$-velocity expectation value $v_{y,n}=\partial E_{n}/\partial k_y$, where $E_n$ are the energy eigenvalues, over all occupied states,  $d I = -e \int dk_y\sum_n v_{y,n}f_n(E_n)$, where $f_n(E_n)$ is the non-equilibrium occupation of the energy eigenvalues. In linear response, the local voltage changes the energy eigenvalues by  $\delta E_n =-e\, dV\int_{x^2+z^2\,<\,r^2}dxdz\,|\psi_n(x,z)|^2$, where $\psi_n(x,z)$ is the wavefunction normalized over the $(x,z)$ plane. Hence, the occupation function changes from the Fermi-Dirac equilibrium distribution $n_F(E_n)$ to $f_n(E_n) = n_F(E_n)+n'_F(E_n)\,\delta E_n$. We set the temperature to zero, in which case the conductance simplifies to
\begin{align}
        G_\mathrm{se}=\frac{e^2}{h}  \sum_{k_y}\mathrm{sign}\big(v_{y,n(k_y)}\big) \mathrlap{\int\limits_{x^2+z^2\,<\,r^2}} \ \ \ \ \ \ \ \ dxdz \, |\psi_{n(k_y)}(x,z)|^2\,, \label{eq.4}
\end{align}
where the sum runs over all $k_y$ at which there is a state $n=n(k_y)$ that crosses the (Fermi) energy, $E_{n(k_y)}=\varepsilon$. We find that the conductance does not depend on the energy $\varepsilon$ in the energy range of valid Weyl-fermion dispersion (see below); at finite temperature $T$ it will thus saturate to $ G_\mathrm{se}$ of Eq.\ \eqref{eq.4} for $k_BT$ lower than that energy range ($k_B $ being the Boltzmann constant).
Figure~\ref{fig:2} shows the step-edge conductance at the upper step edge of size $\nu=1$ as a function of $K$. We find that, independent of energy and system size, the conductance quickly reaches the predicted value~\eqref{ise}, followed by an oscillating tail with a decreasing amplitude $\sim 1/r$. Similar plots are obtained for larger step sizes ($\nu>1$), different energies, and with surface potential (see below and SM \cite{SM} for more data). The values of $r$ at which the conductance is converged does not scale with the system size. Sources of small deviations of the asymptotic values from the expected values are finite-size effects and non-linear corrections to the Weyl-fermion dispersion. The latter can be expected to be more pronounced  at $K=0$ and $K=1$ where the two Weyl nodes merge, explaining the larger deviations there. This upper bound does not scale with system size, addressing one of the open questions posed  by the tilted-surface argument. At very low energies we find deviations due to finite size effects, which vanish with an increasing system size. 

\begin{figure}
    \centering
    \includegraphics[width=\linewidth]{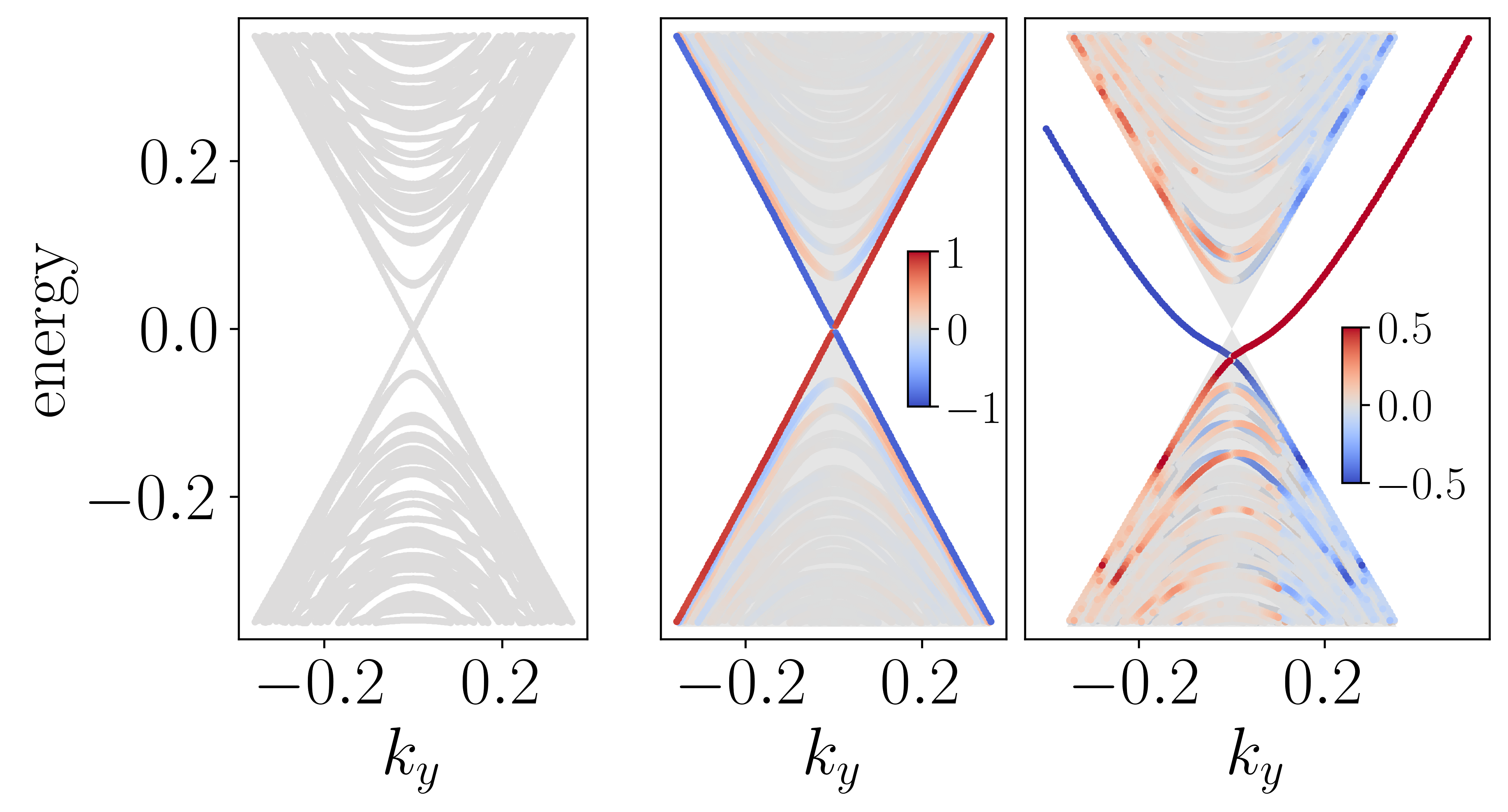}
    \caption{\justifying Dispersion for lattice model for $K=2/3$ and system size of $L_x=L_z=50$ with (left) no step edge, (middle) one pair of step edges without, and (right) with an onsite potential $V=-0.6$ at the step edges. The coloring shows the difference in density of states, with positive (red) and negative (blue) corresponding to larger d.o.s.\ in the top and bottom halves of the system, respectively. The potential splits off one chiral mode at each step and pushes it into the bulk gap, while bulk modes of opposite velocity localize at the step. In gray we color the bulk cone of an infinite system without step edge.} 
    \label{fig:3}
\end{figure}

The numerical results confirm that the conductance assumes the expected value $G_\mathrm{se}= (e^2/h) \, \nu K$. It only depends on the bulk property, namely the separation of the Weyl nodes $K$. It is instructive to see how the non-quantized current is carried by individual states located near the step edge. Instead of fully localized chiral modes, several bulk states have a strongly enhanced weight near the step edge. This can be seen in the $k_y$ dispersion plotted in the middle panel of Fig.\ \ref{fig:3}, where the spatial inhomogeneity of the states is quantified by the probability-density difference between the upper and lower halves of the lattice.  We observe linearly dispersing chiral states confined predominantly to one half of the system, but also additional states at slightly higher energies that are spatially inhomogeneous. Even further inside the cones (both located at $k_y=0$) the states are homogeneous as expected for bulk states. Note also that the non-quantized value can be attained only if the system is large enough that there are sufficient modes that can carry the non-quantized current. Close to the nodal points this is not the case, which is the finite-size effect at low energies mentioned above.

\emph{\clb Effect of a step-edge potential}---While the step edge conductance only depends on the bulk parameter $K$, the distribution of the current among different states and the resulting density of states at the step edge is allowed to depend on surface properties. Indeed, characteristic changes arise if we add a scalar onsite potential on the step-edge sites, the sites on the uppermost and lowermost $z$ layers, next to the step edge. The resulting dispersion is shown in the right panel of Fig.\ \ref{fig:3}. A positive (neg.) scalar on-site potential next to the step edge pushes states localized at the step edge to higher (lower) energies while bulk states remain unaffected. As a result, the localized parts of the states on the lower (upper) dispersion branch split off and form fully localized states within the $k_y$-dependent energy gap. This fully localized in-gap states  contributes $(e^2/h)$ to the conductance, independent of $K$. This alone would violate the predicted step-edge conductance of $(e^2/h)K$. However, at the same energies, bulk states of \emph{opposite velocity} partially localize at the step edge, as can be seen in the plot, so that the total conductance is kept at the universal value of Eq.\ \eqref{ise}, independent of the potential. A qualitatively similar effect occurs if the potential is added to the whole surface (not shown). The potential leads to a strongly enhanced local density of states at energies to one side of the Weyl nodes (see SM \cite{SM} for a quantitative plot). Such a characteristic asymmetric density at step edges around the Weyl nodes agrees with the experimental observation in Ref.\ \cite{Nag2026}.

\begin{figure}
    \includegraphics[width=\linewidth]{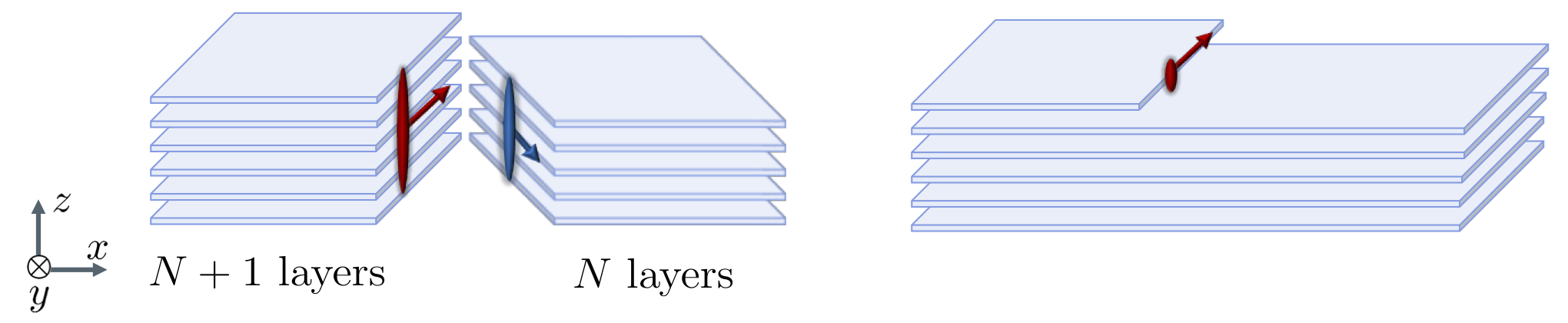}
    \caption{\justifying (Left) Two slabs with $N+1$ and $N$ layers in $z$ direction. (Right) The two slabs coupled into a single slab with a step edge on the top surface.}
    \label{fig:step_from_interface}
\end{figure}

\emph{\clb Analytical Calculation of the Step-Edge Conductance}---As an alternative analytic derivation of the step-edge conductance,  we consider a WSM with $N$ unit cells in the $z$ direction and let the Weyl nodes projected onto the constant-$x$ surface be at $k_z=\pm \pi K$.
The eigenstates are standing waves with quantized momenta $k_{z,l}=l\pi/(N+1),~l=1,\dots,N$. A termination in $x$-direction generates exponentially localized Fermi arc states at $k_{z,l}<\pi K$. The $x$-dependent part of the Fermi-arc wavefunction close to a node, where the Hamiltonian can be expanded to linear order in the $x$-momentum, is of the form
\begin{equation}
    \psi_{k_{z,l}}(x) = \sqrt{2\Delta(k_{z,l})}\, e^{-\Delta(k_{z,l})\, x}\,,\label{wvf}
\end{equation} 
where $\Delta(k_{z,l})$ denotes the inverse localization length, which is proportional to the bulk gap, and the prefactor ensures normalization. We assume a well-defined Weyl semimetal with a bulk gap between the nodes ($k_{z,l}\sim 0$) on the order of $1/K$ and smoothly approaching zero at $k_{z,l} = \pi K$. The current $dI$ induced in response to a surface potential $-e\,dV$ applied in a region of length $r$ from the surface is obtained by summing over all Fermi arc states the probability density integrated in $x$ over the interval~$[0,r]$. For $r\gg 1/K$, almost all Fermi arc states, except those close to the Weyl nodes, are fully localized within the  region, so that the integral is unity for those states, which justifies consideration of wavefunctions close to the nodes, only. The induced current thus reads
\begin{align}
    dI_N &=dV\frac{e^2}{h}\sum_{l\,=\, 1}^{~l_{\textnormal{max}}}\left(1-e^{-2\Delta(k_{z,l}) r}\right)\,,\label{singleslab}
\end{align}
$k_{z,l_{\textnormal{max}}}$ being the largest momentum satisfying $k_{z,l}<\pi K$. We now consider two copies of this finite slab with $N+1$ and $N$ layers along~$z$, respectively, placed next to each other such that one of the slabs has one more layer on the top, as shown in Fig.\ \ref{fig:step_from_interface}.
When uncoupled, the two slabs give rise to the chiral surface currents $dI_{N+1}$ and  $-dI_N$ at the interface. A straight-forward calculation of the total interface current $dI\equiv dI_{N+1}-dI_N$ under the assumption $r\ll N$ leads to
\begin{equation}
   dI =  dV\, (e^2/h)\,K
\end{equation}
(see SM \cite{SM} for a detailed calculation). To connect this result with the current carried by a step edge, 
we consider what happens upon introducing a local hopping at the interface between the two slabs, which would merge the two slabs into a single slab with a step edge on the top surface as displayed in Fig.\ \ref{fig:step_from_interface}.
While the two interface surfaces are modified by the introduced hopping, the total current response $d I$ is unchanged. This can be seen by considering that application of $dV$ on the outer opposite surfaces of the two slabs would compensate the interface-induced current $dI$. If the local hopping at the interface would modify $dI$ it would modify the response at the outer surfaces as well,  which contradicts the locality of the hopping-amplitude change. The interface hopping can thus be chosen such that the interface disappears and the upper and lower surfaces become featureless, except for the step-edge on the upper surface. Since the featureless surfaces do not support a current response in the $y$ direction, the current response $d I$ must be uniquely associated with the step edge.  Note that this analytic calculation of the step-edge conductance is valid also at energy $\varepsilon=0$, which was inaccessible numerically due to the finite size.

\begin{figure}[t]
    \centering
    \includegraphics[width=\linewidth,clip=true]{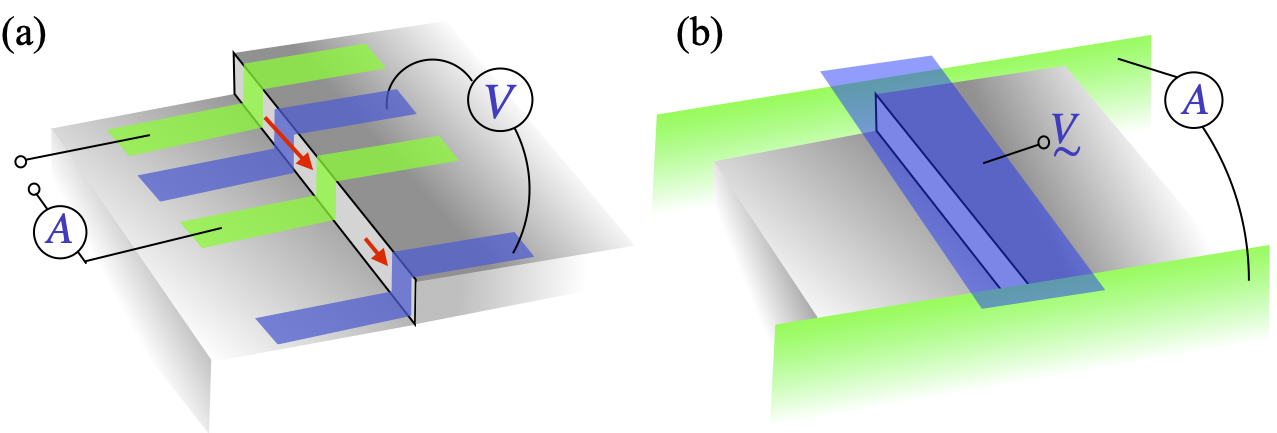}
    \caption{\justifying Schematics of possible experimental setups to detect chiral step-edge conductance. (a) Four-terminal setup. (b) Three-terminal setup with ac voltage bias.}
    \label{fig:4}
\end{figure}

\emph{\clb Discussion}---The non-integer step-edge conductance $G_\mathrm{se}=(e^2/h) \,(\nu |\bm{K}\times \bm{e}_\mathrm{se} |)$ is universally determined by the separation of a pair of Weyl nodes in the bulk and the step-edge orientation. In case of multiple Weyl-node pairs and additional non-trivial bulk topology, all contributions add up independently, whereby the non-integer part of $G_{\rm se}$ related to the positions of the Weyl node pairs, derived here, can be supplemented with an integer part, associated with the topology characteristic of a three-dimensional (stacked) Chern insulator.

Although the experimental verification of the step-edge conductance requires transport measurements, some qualitative properties of the local density of states detected in \cite{Nag2026} are consistent with our theory: The density of states is observed to have an asymmetric energy dependence around the Weyl nodes, which we identified as a hallmark of a chiral state separated from the bulk continuum by a generic local potential at the step edge. Furthermore, the density of states along the step edge observed in \cite{Nag2026} remains remarkably smooth despite evident surface disorder, which one can understand as prohibited standing-wave formation when the weights of forward- and backmoving wavefunctions are different. In the presence of disorder, we expect that the chiral step-edge currents enjoy the same robustness as the currents carried by Fermi arcs on extended  disordered surfaces of topological metals \cite{Wilson2018}. When interactions become strong enough to overcome topological protection of Fermi arc states, we expect the step-edge conductance to lose its robustness as well \cite{interactions1, interactions2, interactions3}.

Interesting for future experiments is a direct detection of the step-edge anomaly. This can be achieved in magnetic topological metals in a four-terminal setup, measuring a voltage difference between two parts of the step edge, one of which conducts a current, illustrated in Fig.\ \ref{fig:4}(a). A quantitative measurement of the non-integer step-edge conductance can be achieved by applying a local ac voltage bias along the step and collecting the resulting current flow by leads, as illustrated in Fig.\ \ref{fig:4}(b). Such experimental probes of Weyl topology are especially valuable in systems where opposite-chirality Weyl nodes project onto one another on the natural surfaces, such as (Cr,Bi)$_2$Te$_3$ \cite{TookaWeyl}, EuCd$_2$As$_2$ \cite{idealWSM, SinglePairWeylnodes}, and MnBi$_2$Te$_4$ \cite{IdealWSM2}. In these materials, the overlapping surface projections preclude the conventional surface signatures of Weyl topology, making the step-edge signature proposed here a unique manifestation of the underlying topological phase.

In closing, it is interesting to point out that the robust step-edge conductance could play an important role in the mechanism of record-low resistivity measured in topological-metal nano-structures  \cite{Zhang2019,Cheon2025}. Their  boundaries naturally feature step edges that, depending on the sample geometry and the density of step edges, could dominate the transport. Generally, in combination with emerging microstructuring capabilities \cite{Seo2026}, our discovery of one-dimensional chiral transport channels on surfaces of stable three-dimensional matter opens a new route for harnessing robust transport in quantum electronics

\emph{\clb Data Availability}---All the code and data used to produce the reported results is available in Ref.\ \cite{Schweizer}.

\emph{\clb Acknowledgments}---We thank Nurit Avraham, Haim Beidenkopf, Ching-Tzu Chen, and Ady Stern for useful discussions. This work was supported by the Deutsche Forschungsgemeinschaft (DFG, German Research Foundation) - Project Number 277101999 - CRC-TR 183 (projects A02, A03),  the Emmy Noether program - Project Number 506208038, the European Research Council (ERC) Consolidator grant No. 101042707 (topomorph) and Cluster of Excellence Exc 3112 Center for Chiral Electronics.
\nocite{NotWeyl,Groth_2014}
\bibliography{library}
\newpage
\setcounter{equation}{0}
\renewcommand{\theequation}{S\arabic{equation}}
\renewcommand{\thefigure}{S\arabic{figure}}
\newpage
\onecolumngrid
\newpage
\section*{Supplemental Material}

\section{System-size and energy dependence of step-edge conductance}\label{appendixA}
\begin{figure}[h]
    \begin{subfigure}{0.3\linewidth}
        \includegraphics[width=\linewidth,clip=false]{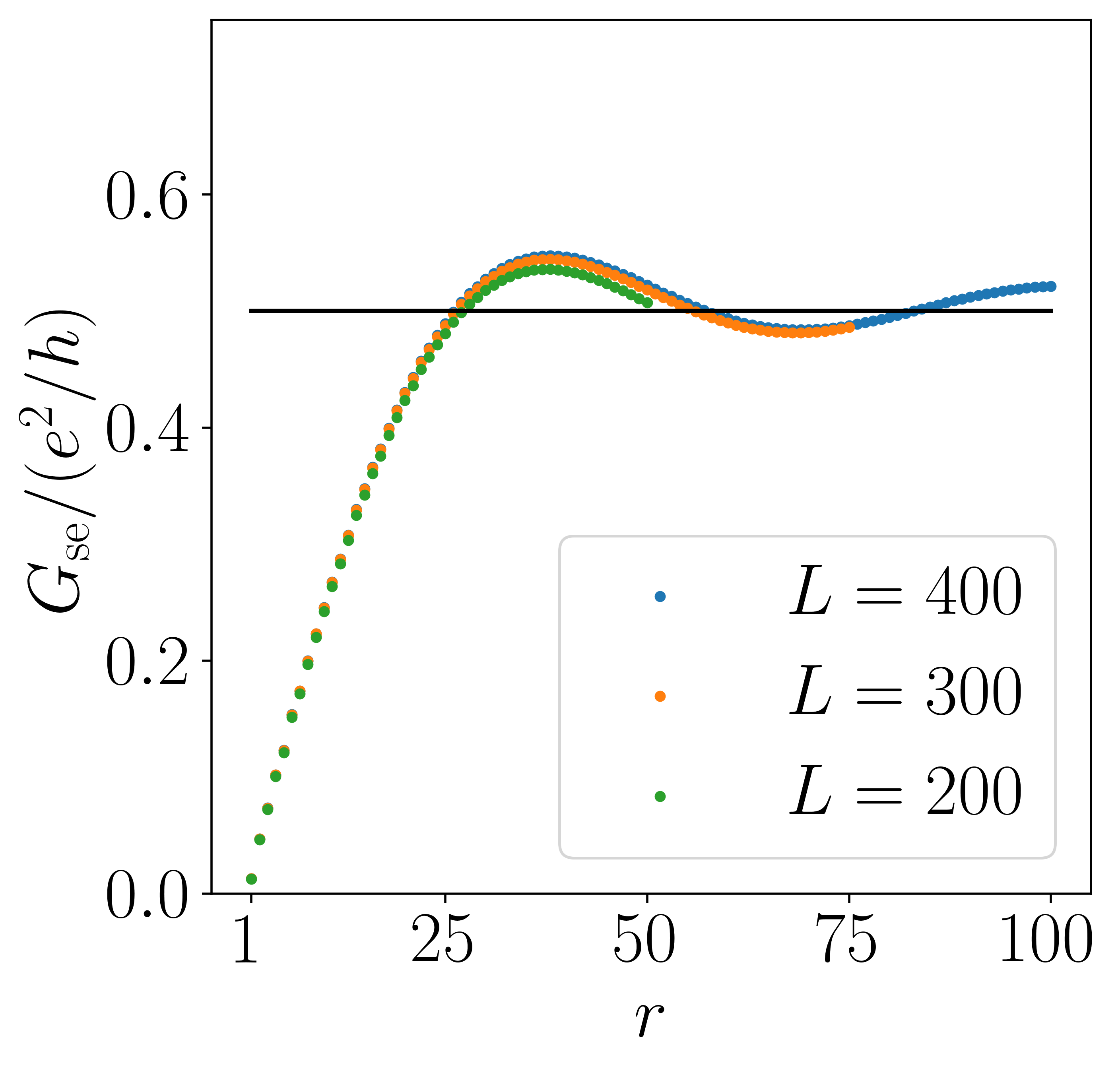}
        \caption{\centering }
        \label{fig:A1}
    \end{subfigure}
    \hfill
    \begin{subfigure}{0.3\linewidth}
        \includegraphics[width=\linewidth,clip=false]{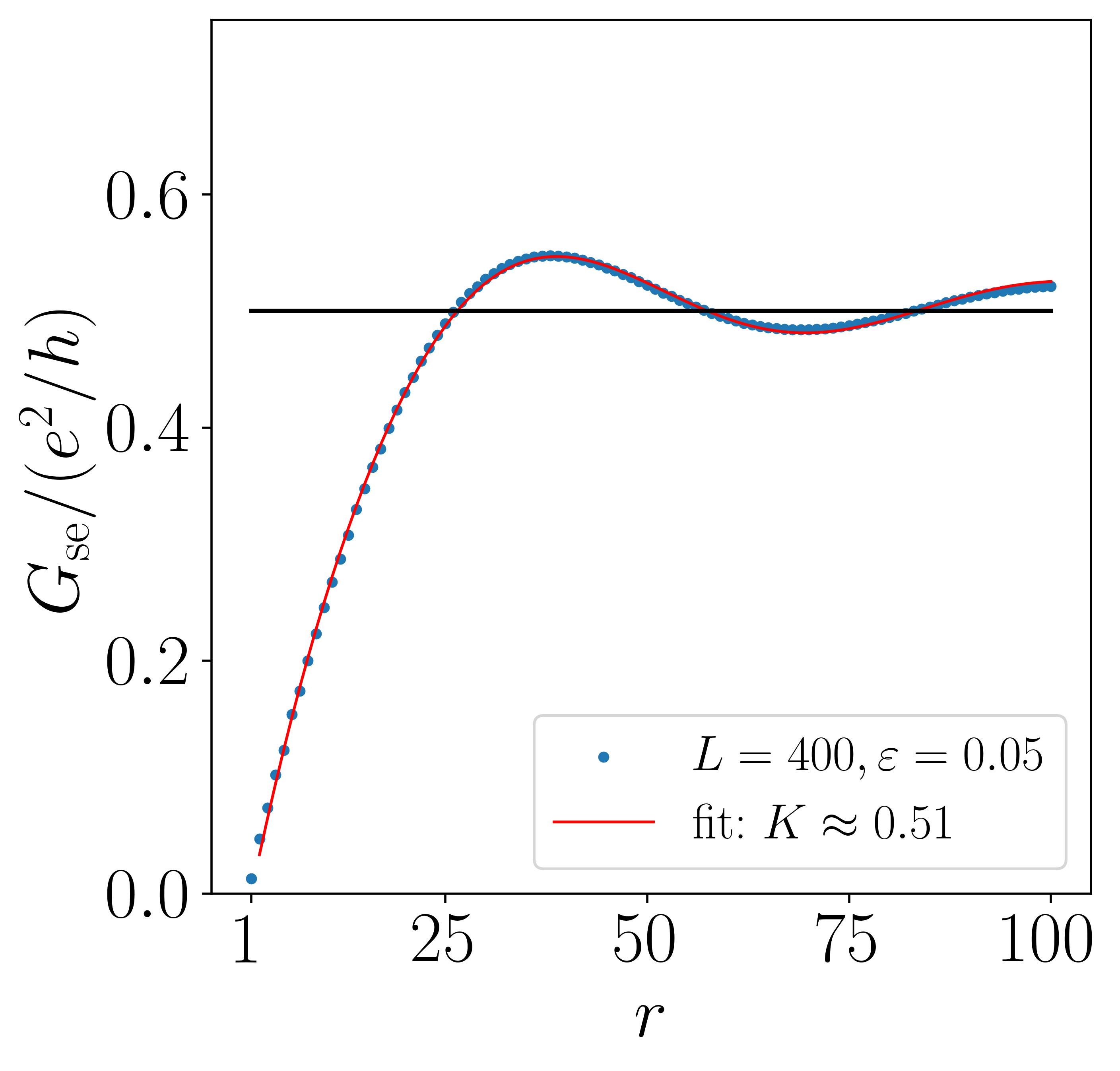}
        \caption{\centering}
        \label{fig:A2}
    \end{subfigure}
    \hfill
    \begin{subfigure}{0.3\linewidth}
        \includegraphics[width=\linewidth,clip=false]{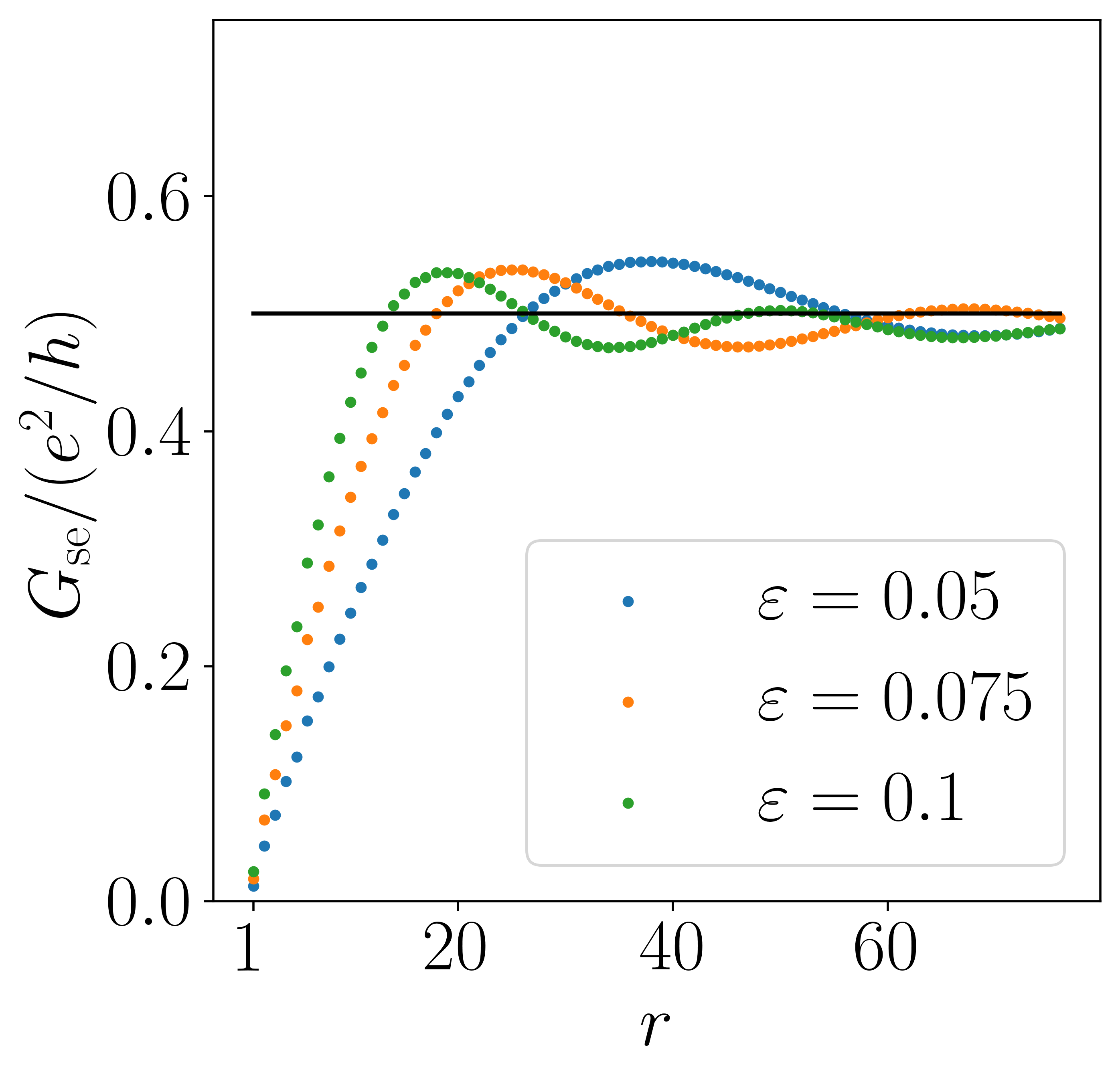}
        \caption{\centering }
        \label{fig:A3}
    \end{subfigure}
    \caption{\justifying Step-edge conductance $G_{\mathrm{se}}/(e^2/h)$ as a function of the radius of the half circle integration region at $K=1/2$.  The black line is prediction of Eq.\ (2) of the main text. (a) $G_{se}/(e^2/h)$ for different $L_x=L_z=L$ at $\varepsilon=0.05$. The radius is restricted to $25\%$ of the system size. (b) $G_{\mathrm{se}}/(e^2/h)$ for system size $L=400$ at $\varepsilon=0.05$ and $K=1/2$ together with the fit to the function $f(r)$ as described in the text. (c) $G_{\mathrm{se}}/(e^2/h)$ for different energies for system size $L_x=L_z=300$ }
    \label{fig:A1all}
\end{figure}

In this section we examine the dependence of the step-edge conductance on system size, demonstrate how we extract the asymptotic value for large integration region, and validate the choice of the energy. Figure \ref{fig:A1all} shows the numerical evaluation of the step-edge conductance $G_{\mathrm{se}}$ 
\begin{align}
        G_\mathrm{se}=\frac{e^2}{h}  \sum_{k_y}\mathrm{sign}\big(v_{y,n(k_y)}\big) \mathrlap{\int\limits_{x^2+z^2\,<\,r^2}} \ \ \ \ \ \ \ \ dxdz \, |\psi_{n(k_y)}(x,z)|^2,
\end{align}
as a function of radius $r$ of the integration half circle, where the sum runs over all $k_y$ at which there is a state $n=n(k_y)$ that crosses the (Fermi) energy, $E_{n(k_y)}=\varepsilon$. In Fig.\ \ref{fig:A1all}(a) we show results for different system sizes $L_x=L_z=L\in \left\{200,300,400\right\}$ at fixed energy $\varepsilon=0.05$. The amplitude and frequency of oscillations is independent of system size, showing quantitatively that the localization of the step-edge response is an intrinsic property. Larger system sizes allow to resolve larger radii making it easier to extract the value of convergence. Still we do not fully reach convergence within system sizes that we can achieve in our numerical calculations, so we obtain the convergence value with a fit. The extraction of convergence value is exemplified in Fig.\ \ref{fig:A1all}(b), in which we fit the function $f(r)=A\,\left(1-\exp(B\,r)\right)+C\,\sin\left(D\,(r-r_0)\right)/(r-r_0)$ to the conductance. This choice of $f(r)$ allows to find the converging value with the overlying oscillations of decreasing amplitude. For minimally oscillating values of $K$ we set the parameter $C=0$. Fit parameters are $A, B, C, D, r_0$ where $A$ corresponds to the asymptotic value. To quantitatively show independence of our results on the choice of energy $\varepsilon$, Fig. \ref{fig:A1all}(c) shows the step-edge conductance for different energies. Although the oscillation frequencies have a strong dependence on $\varepsilon$, the choice of energy does not significantly change the oscillation amplitude nor the asymptotic value. At the energy of $\varepsilon=0.05$ for $L_x=L_z=L=400$ we are far away from the finite energy gap.

\section{Enhanced density of states at the step edge due to surface potential}\label{appendixB}
\begin{figure}[h]
    \begin{subfigure}{0.49\linewidth}
        \includegraphics[height=5.5cm]{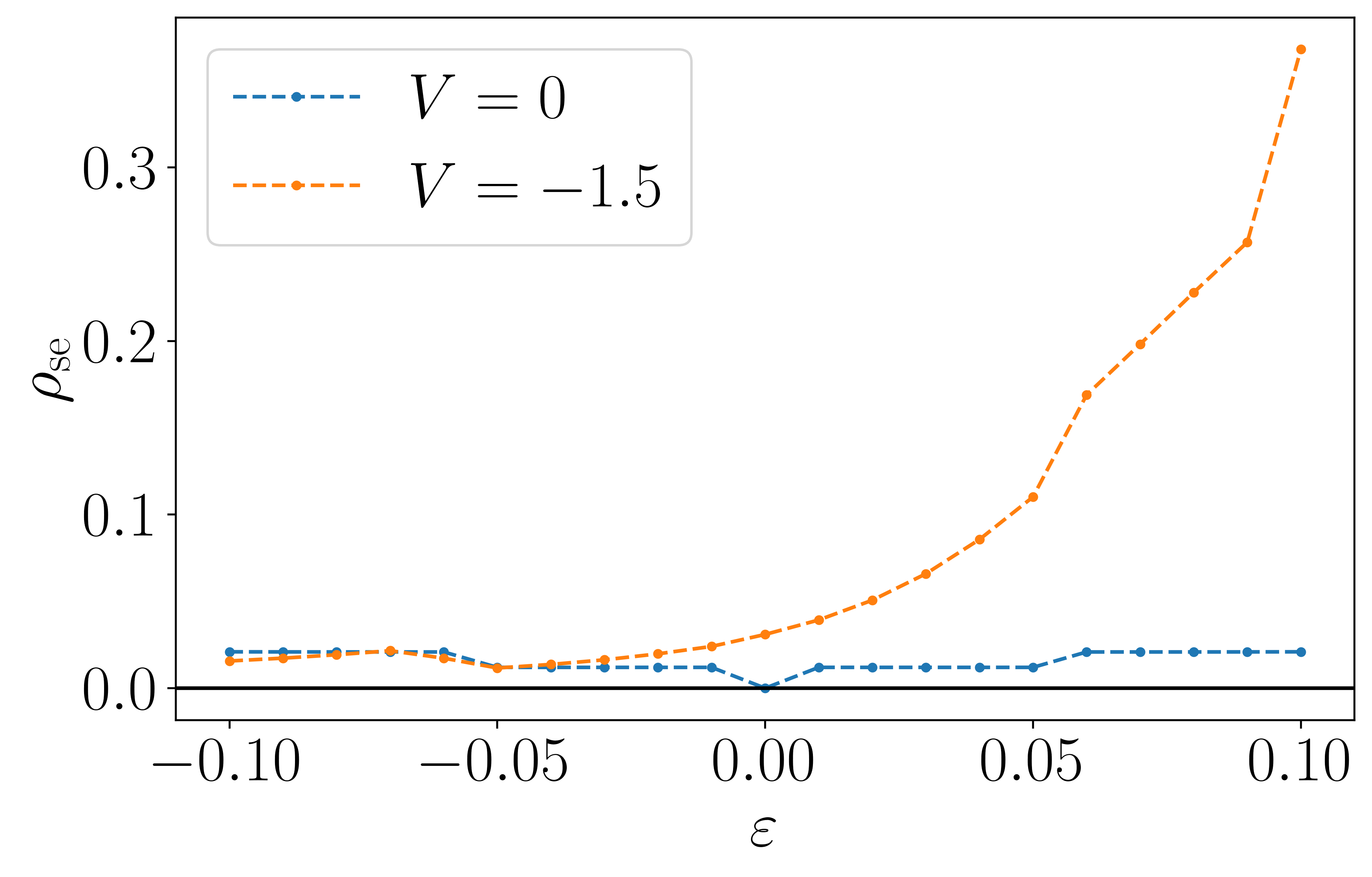}
        \caption{\centering}
        \label{A4.1}
    \end{subfigure}
    \hfill
    \begin{subfigure}{0.245\linewidth}
        \includegraphics[height=5.5cm]{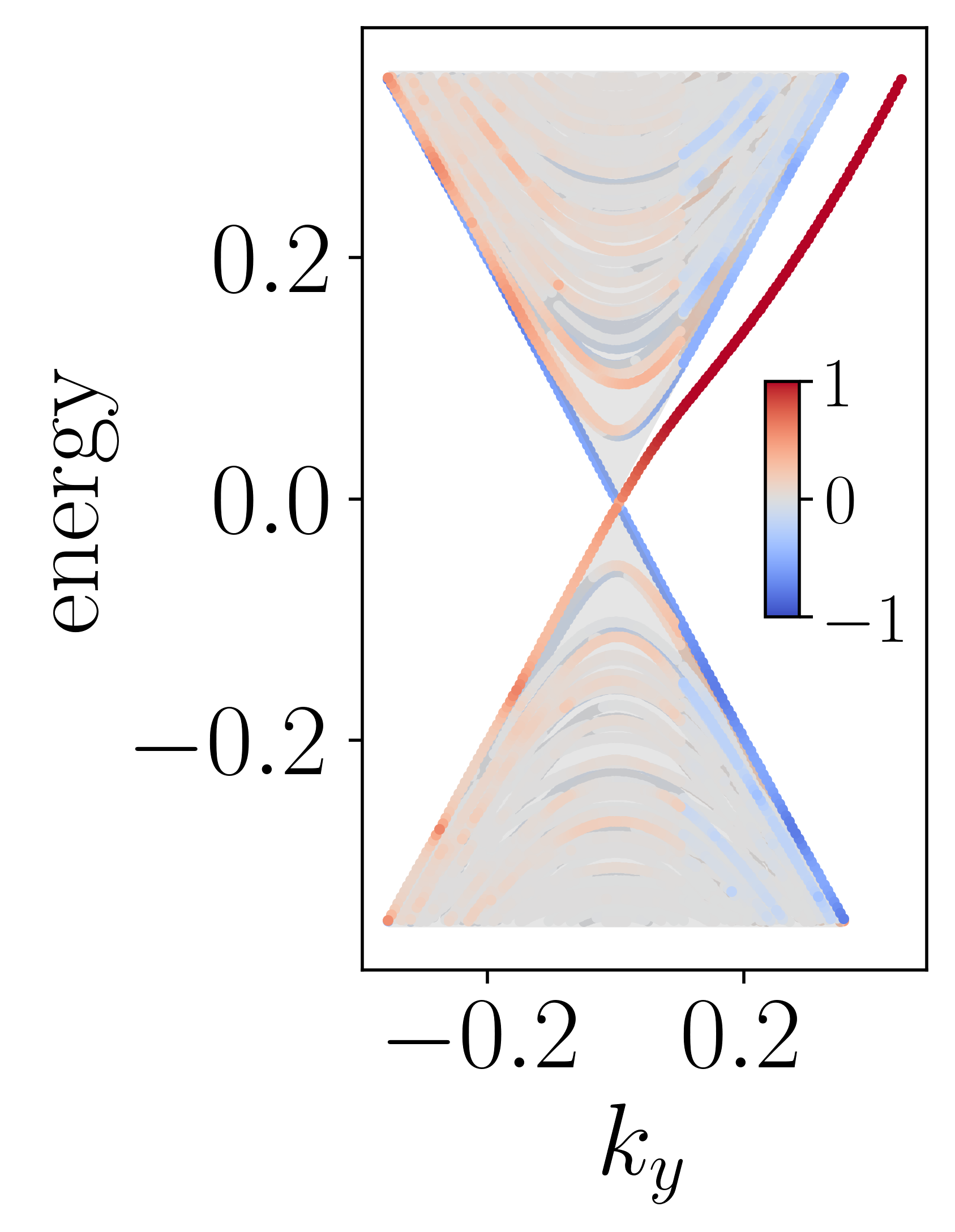}
        \caption{\centering }
        \label{A4.2}
    \end{subfigure}
        \hfill
   \begin{subfigure}{0.245\linewidth}
        \includegraphics[height=5.5cm]{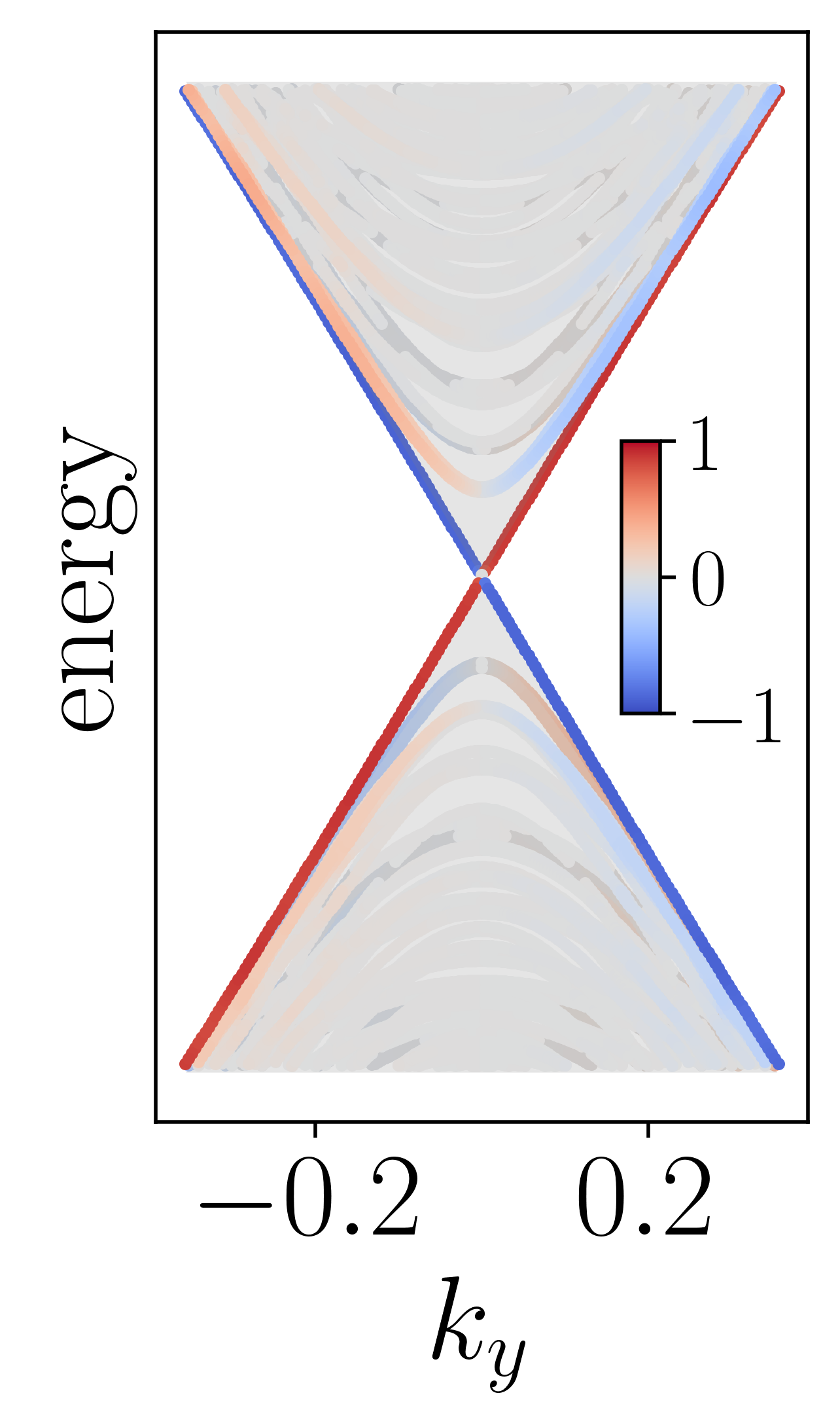}
        \caption{\centering }
        \label{A4.3}
    \end{subfigure}
    \caption{\justifying (a) Local density of states at the step edge $\rho_{\mathrm{se}}$ for $L_x=L_z=50$ and $K=1/3$. Without a potential on the step edge, the states contributing to the chiral current also contribute to the density of states exactly on the step edge. In this case, the density of states is symmetric around zero energy as for the model of Eq.\ (3) of the main text. When adding a negative scalar potential on the step edge, a localized state in the $k_y$-dependent bulk energy gap emerges at the edge. This leads to a strongly enhanced density of states at the step edge for energies $\varepsilon > 0$, as shown by the red curve in panel (a). (b) Dispersion  for $K=1/3$, system size $L_x=L_z=50$, and $V=-1.5$. (c) Dispersion  for $K=1/3$, system size $L_x=L_z=50$, and $V=0$}
    \label{fig:A4}
\end{figure}

This appendix shows the enhanced density of states at the step edge when a negative scalar potential is applied precisely at the top of the step edge, that is to say, the unit cell with which the top layer terminates. Figure \ref{fig:A4}(a) displays the local density of states close to the step edge. We probe this by projecting the density of states onto the three unit cells that make up the step edge as a function of energy
\begin{equation}
    \rho_{\mathrm{se}}=\sum_{k_y}\  \mathrlap{\int\limits_{x^2+z^2\,<\,1}} \ \ \ \ \ \ \ \ dxdz \, |\psi_{n(k_y)}(x,z)|^2,\label{eq_S2}
\end{equation} 
where the sum runs over all $k_y$ at which there is a state $n=n(k_y)$ at energy $\varepsilon$. The strongly enhanced local density of states at the step edge for $\varepsilon > 0$ in the presence of the negative edge potential shown in Fig.\ \ref{fig:A4}(a) is attributed to the formation of a localized edge mode in the $k_y$-dependent gap for positive energies, see Fig.\ \ref{fig:A4}(b). For reference, Fig.\ \ref{fig:A4}(c) shows the $k_y$-dependent spectrum without edge potential. 

\section{Analytical calculation of the step-edge conductance}\label{appendixC}

In this section we analytically derive the step-edge conductance. To that end we will consider a step edge geometry arising from coupling two Weyl semimetals (WSMs) with $N$ and $N+1$ layers in $z$-direction as outlined in the main text. We start from a single WSM slab of $N$ discrete layers in the $z$ direction, semi-infinite in $x$, with a surface at $x=0$, and infinite in $y$. The Weyl nodes project onto the $x=0$ surface at $k_z=\pm \pi K$, $k_y=0$. The eigenstates are standing waves in $z$, with quantized momenta $k_{z,l}=l\pi/(N+1),~l=1,\dots,N$. At finite energy $\varepsilon=v k_F$, with $v$ the velocity of the Weyl cone, the systems surface at $x=0$ without a surface potential hosts straight Fermi arcs at momenta $|k_{z,l}|<\pi K$, $k_y=k_F$. Close to the nodes, where the infinite-system Hamiltonian can be linearized in momentum $k_x$ measured from the node, the $x$-dependent part of the Fermi-arc wavefunction is  exponentially decaying, with the decay length inversely proportional to the bulk gap at the Fermi arc momenta, which is a function of $k_{z,l}$ denoted as $\Delta(k_{z,l})$, hence 
\begin{equation}
    \psi_{k_{z,l}}(x) = \sqrt{2\Delta(k_{z,l})}\, e^{-\Delta(k_{z,l}) x}\,,\label{Appwvf}
\end{equation} 
where the prefactor ensures normalization. It is sufficient to assume that the inverse decay length vanishes at $k_z=\pi K$ (where the Fermi arc merges with the bulk cone), and smoothly increases to $\sim1/K$ towards the middle of the arc, i.e., $\Delta(0)\sim1/K$.

We now consider two WSMs with identical bulk dispersion, with one consisting of $N$ layers in $z$-direction and extending semi-infinitely in negative $x$ and the second consisting of $N+1$ layers in $z$ and extending semi-infinitely in positive $x$. They share a surface at $x=0$, such that when the two systems are coupled and merge into one system, a step edge forms on the top surface. Before coupling, the $x=0$ surface host chiral surface currents $-dI_N$ and $dI_{N +1}$ in response to a voltage $dV$ applied in the region $|x|<r$. The conductance $(dI_{N +1} - dI_{N})/dV$ of the interface is thus given by
\begin{align}
  \frac{dI_{N +1} - dI_{N}}{d V\, e^2/h} =&\  \sum_{l\,=\, 1}^{l_{\mathrm{max}}}\left(1-e^{-2\Delta(k_z)r}\right) -\sum_{l\,=\, 1}^{l_{\mathrm{max}}'}\left(1-e^{-2\Delta(k_{z,l}')r}\right)\,,\label{ab1}
\end{align}
where the first sum corresponds to the left surface consisting of $N+1$ layers with $z$-momenta quantized as \mbox{$k_{z,l}=l\pi/(N+2)$} and the right sum corresponds to the right surface consisting of $N$ layers with $z$-momenta quantized as \mbox{$k_{z,l}'=l\pi/(N+1)=k_{z,l}+\delta k_{z,l}$, where $\delta k_{z,l}=l\pi/((N+1)(N+2))=k_{z,l}/(N+1)$}. Individually, both sums diverge as $N\rightarrow\infty$ due to the increasing number of surface modes, yet calculating their sum before taking the limit shows convergence to a finite value, which corresponds to the localized current at the step edge.\\ 
We calculate this current in the \mbox{$KN\gg Kr \gg 1$} limit and drop terms of order $\mathcal{O}\left(r/N\right)$ and $\mathcal{O}\left(1/Kr\right)$. This means that the bulk gap function on the right surface is evaluated at $\Delta(k_{z,l}+\delta k_{z,l})\approx\Delta(k_{z,l})+\delta k_{z,l}\Delta'(k_{z,l})$. The maximal values $n$ and $n'$ in both sums may be equal ($l_{\textnormal{max}}=l_{\textnormal{max}}'$) or differ by 1 ($l_{\textnormal{max}}=l_{\textnormal{max}}'+1$) as it is not apriori known whether the differently quantized $z$-momenta produce an additional Fermi arc state or not. Let us first assume no additional Fermi arc state is present ($l_{\textnormal{max}}=l_{\textnormal{max}}'$). Noting that the summation index $l$ can be of order $N$ we get:
\begin{align*}
     \frac{dI_{N +1} - dI_{N}}{dV\, e^2/h} &=  \sum_{l\,=\, 1}^{l_{\mathrm{max}}}\left(-e^{-2\Delta(k_{z,l})r}+e^{-2\Delta(k_{z,l} + \delta k_{z,l})r}\right)\,,
\end{align*}
where we can readily expand to linear order in $ \delta k_{z,l}r  \sim r/N$: $e^{-2\Delta(k_{z,l} + \delta k_{z,l})r}=e^{-2\Delta(k_{z,l})r}\left(1 - 2\delta k_{z,l}\Delta'(k_{z,l}) r\right)$ to obtain
\begin{align}
     \frac{dI_{N +1} - dI_{N}}{d V\, e^2/h} =-2\sum_{l\,=\, 1}^{l_{max}}\delta k_{z,l}\Delta'(k_{z,l})r\, e^{-2\Delta(k_{z,l})L}=&\, -\frac{2r}{N+1}\sum_{l\,=\, 1}^{l_{max}}k_{z,l}\Delta'(k_{z,l})\, e^{-2\Delta(k_{z,l})r} \nonumber \\
     &\approx- \frac{2r}{\pi}\int_0^{\pi K} k_{z} \Delta'(k_{z}) \, e^{-2\Delta(k_{z})r} dk_{z}\,.
\end{align}
Where we use the limit of $N\rightarrow\infty$ to go from discrete $k_{z,l}$ to continuous momenta $k_{z}$ and rewrite the sum as an integral. Integration by parts gives
\begin{align*}
     \frac{dI_{N +1} - dI_{N}}{dV\, e^2/h} &= \frac{1}{\pi} \left[ k_{z} \, e^{-2\Delta(k_{z})r} \right]_0^{\pi K} - \frac{1}{\pi} \int_0^{\pi K}  e^{-2\Delta(k_{z})r} dk_{z} = K e^{-2\Delta(\pi K)r} - \frac{1}{\pi} \int_0^{\pi K}  e^{-2\Delta(k_{z})r} dk_{z}\,.
\end{align*}
Since the gap vanishes at $k_z=\pi K$  and smoothly increases to $\sim1/K$ towards the middle of the arc, the remaining integral is on the order $\mathcal{O}\left(1/rK\right)$ and the step-edge conductance is predicted to be
\begin{equation}
    \frac{G_{\mathrm{se}}}{(e^2/h)}=K+ \mathcal{O}\left(\frac{1}{rK}\right)\,. \label{Aeq4}
\end{equation}
Given the other case of $l_{\textnormal{max}}=l_{\textnormal{max}}'+1$, the larger surface has an additional contribution from the extra Fermi arc state,
\begin{align*}
    1- e^{-2\Delta(\pi K - k_{z,l_{\textnormal{max}}+1})r}\simeq 2(\pi K - k_{z,l_{\textnormal{max}}+1})\Delta'(\pi K)r\,.
\end{align*} 
Considering the spacing of quantized momenta is $\sim1/N$ and $k_{z,l_{\textnormal{max}}+1}$ is the largest $z$-momentum still on the Fermi arc, this additional contribution goes as $r/N$ and is negligible in the $N\gg r$ limit, and thus yields the same step-edge conductance as in \eqref{Aeq4}.

\section{Transfer matrix method}\label{appendixD}
In order to calculate the current response at a given energy we solve for the propagating modes in $y$-direction at fixed energy $\varepsilon$ of a three-dimensional WSM. To find the propagating wavefunctions we use a transfer matrix approach. The system is translationally invariant in $y$, finite in $x$ and $z$ with periodic and open boundary conditions, respectively. For $\psi_{n}$ being the wavefunction in $xz$-layer $n$, $H_l$ the Hamiltonian of a single $xz$-layer and $T$ the hopping to the next layer in $y$-direction, the tight-binding formulation of the dynamics in $k_y$ are described by
\begin{align}
    \varepsilon \psi_{n}=T\psi_{n+1}+H_{l}\psi_n+T^{\dagger}\psi_{n-1}\,.
\end{align}
For non-singular $T$ equivalently 
\begin{align}
    \psi_{n+1}=T^{-1}(\varepsilon\mathds{1}-H_{l})\psi_n-T^{-1}T^{\dagger}\psi_{n-1}\,,
\end{align}
which can be written into a matrix equation defining the transfer matrix $\mathcal{T}$:
\begin{align}
\begin{pmatrix}\psi_{n+1}\\\psi_n\end{pmatrix} &= \begin{pmatrix} T^{-1}(\varepsilon\mathds{1}-H_{l}) & -T^{-1}T^{\dagger} \\ 1 & 0\end{pmatrix}\begin{pmatrix}\psi_{n}\\\psi_{n-1}\end{pmatrix} \equiv \mathcal{T} \begin{pmatrix}\psi_{n}\\\psi_{n-1}\end{pmatrix}\,.
\end{align}
For our lattice model $T(k_y)=\mathds{1}_l\otimes[\cos(k_y)\sigma_z+\sin(k_y)\sigma_y]$ describes the hopping in $y$-direction and the layer Hamiltonian is $H_l(k_x, k_z) = [2-\cos(k_x) + \cos(\pi K)-\cos(k_z)]\, \sigma_z + \sin(k_x) \, \sigma_x$. The tight-binding Hamiltonian is constructed using the \textit{Kwant} package \cite{Groth_2014}. We solve for the eigenvalues $\lambda_n$ of the transfer matrix $\mathcal{T}$ at fixed energy $\varepsilon$. The solutions of $|\lambda_n|=1$ are propagating states with momenta $k_{y,n}=\mathrm{arg}(\lambda_n)$.
\end{document}